# The Missing Baryons in the Milky Way and Local Group

A White Paper submitted to ***The Galactic Neighborhood*** Science Frontiers Panel

Joel N. Bregman
Department of Astronomy
University of Michigan
Ann Arbor, MI 48109-1042
Email: jbregman@umich.edu
Telephone: 734-764-3454

Robert A. Benjamin: University of Wisconsin - Whitewater
Massimiliano Bonamente: University of Alabama in Huntsville
Claude R. Canizares, Massachusettes Institute of Technology
Ann Hornschemeier: NASA Goddard Space Flight Center
Edward Jenkins: Princeton University
Felix J. Lockman: National Radio Astronomy Observatory, Green Bank
Fabrizio Nicastro: Harvard Smithsonian Center for Astrophysics
Takaya Ohashi: Tokyo Metropolitan University
Frtis Paerels: Columbia University
Mary E. Putman: Columbia University
Kenneth Sembach: Space Telescope Science Institute
Norbert Schulz: Massachusettes Institute of Technology
Blair Savage: University of Wisconsin
Randall Smith: Harvard Smithsonian Center for Astrophysics
Steve Snowden: NASA/GSFC
Noriko Yamasaki: ISAS/JAXA
Yangsen Yao: University of Colorado
Bart Wakker: University of Wisconsin

## 1. The Science Issues

The Milky Way and all other galaxies are missing most of their baryons in that the ratio of the known baryonic mass within $R_{virial}$ to the gravitating mass, $\Omega_{baryon}/\Omega_M$ is 3-10 times less than the cosmic ratio determined from WMAP (e.g., Hoekstra et al. 2005, McGaugh 2007). This is verified not only in studies of other individual galaxies (e.g., M33 is missing 90% of its baryons; Corbelli 2003) but in large ensembles of galaxies, based on weak lensing studies (e.g., Gavazzi et al. 2007; Parker et al. 2007). Evidently, galaxy formation and evolution leads to a separation between baryons and dark matter. Not only is the energy needed to produce this separation enormous, it must be central to the galaxy formation process.

One possibility is that the baryons fell into the dark matter potential and were subsequently ejected during a galactic wind phase, driven by supernovae or an active galactic nucleus. The other possibility is that most of the gas never collapsed into the dark matter potential because it was heated prior to the collapse. This pre-heating would be spatially distributed and most likely due to supernovae, such as during a Population III phase. In either case, understanding the flow of baryons into and out of galaxies in the evolving universe is essential for ultimately understanding galaxy formation and evolution.

Low redshift galaxies and their environments are the end result of these heating and mass inflow/outflow events over cosmic time. Fortunately, the events producing the low redshift universe leave distinctive observational signatures. For example, the overall heating and ejection (or pre-heating, with little ejection) could be deduced from the extent and temperature of the gas today, along with the mass of metals produced and their elemental ratios. To answer these scientific questions, we must

- ***Discover the missing Galactic and Local Group baryons by measuring the extended distribution of baryons around the Milky Way and in the Local Group.***

- ***Measure the temperature and metallicity properties of the missing baryons to determine the processes responsible for the baryon/dark matter separation***

The characteristic temperature of a potential well like the Milky Way is 1-3$\times10^6$ K, so it is critical to trace this hot gas, which is only detectable in X-rays. Perhaps the best place to discover and study missing galactic baryons is in the vicinity of the Milky Way, where both absorption and emission studies can be brought to bear.

Surrounding the Milky Way is a multiphase halo of gas that is a few kpc in thickness. The hot component was detected with *ROSAT*, where it is found to have a temperature of $1\times10^6$ K (e.g., Pietz et al 1998; Snowden et al. 2000) and a scale height of about 4 kpc, which is similar to hot halos seen around other spirals. Also, there is considerable OVI gas (Bowen et al. 2008), indicative of gas at $10^{5.5}$ K if it is in ionization equilibrium. Lower ionization absorption line gas is also observed, as is neutral hydrogen. This picture is consistent with material cooling from $\sim10^6$ K, such as in a galactic fountain or galactic accretion.

Another hot component, a larger halo was discovered in the last few years through the detection of X-ray absorption lines (OVII, OVIII; Nicastro et al. 2002; Rasmussen, Kahn, and Paerels 2003). As traced by the OVII absorption line (Figure 1), a $10^6$ K gas diagnostic, the extent of this hot material can be estimated, and although poorly known, it is likely to be in the range 20-100 kpc (Bregman and Lloyd-Davies 2007; Yao et al. 2008). This is similar to the size of the high velocity clouds of HI around the Milky Way and M31, which lie within 50 kpc of their host system. None of these components, hot or cold, account for the missing baryons.

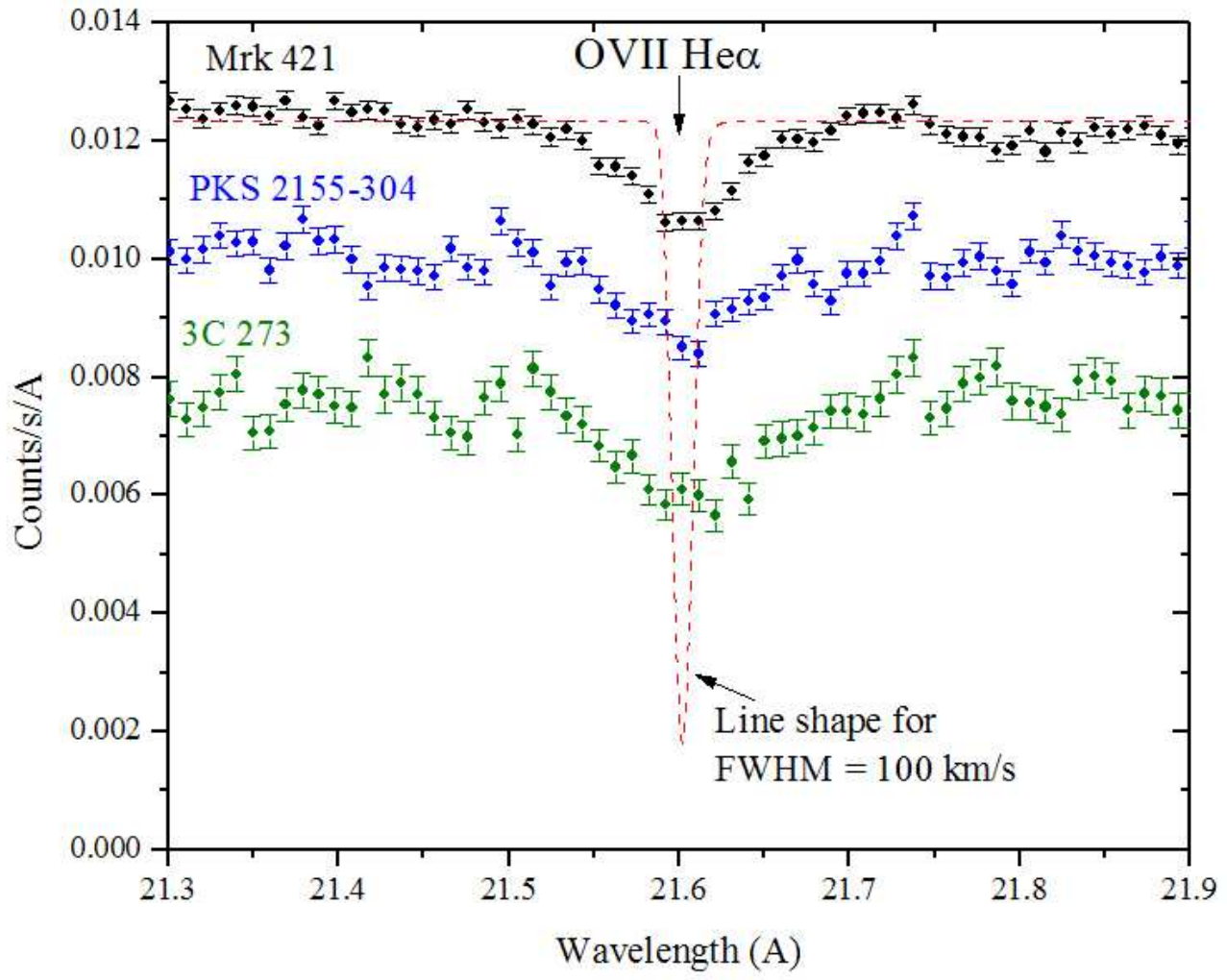

**Figure 1.** The three brightest AGNs show absorption from OVII Heα at $z \approx 0$, indicative of $10^6$ K gas around the Milky Way or in the Local Group. The instrumental resolution of these *XMM RGS* spectra is about 360, only slightly worse than that of *Chandra*, but far less than the R = 3000 value that is needed to resolve the expected line shape (in red).

There are clues that a more extensive hot medium is present. The high velocity clouds of HI, at a distance of 5-20 kpc from the disk, must be interacting with a hotter more dilute medium to produce the ubiquitous OVI absorption features (Fox et al. 2004). A hot dilute medium on a scale of a few hundred kpc is inferred from the apparent stripping of HI in Local Group dwarf galaxies (Blitz and Robishaw 2000; Grcevich and Putman 2009). They find that for dwarfs within 270 kpc of the Milky Way or M31, HI is not detected, but it is present in the more distant dwarfs. From this, the mass of the ambient medium responsible for stripping is inferred. This dilute and presumably hot medium represents $\sim 10^{10}$ $M_{\odot}$, which is impressive, but still not sufficient to account for the missing galactic baryons. However, if this medium was comparable to the size of the Local Group or larger, the gas mass would be of cosmological significance.

The evidence points to a Milky Way halo plus a more extensive Local Group medium, so to characterize the properties of each, they must be separated observationally. There are three ways in which these two components can be distinguished observationally: by temperature; by velocity; and by spatial distribution on the sky. In terms of temperature, the characteristic potential well depth of the Milky Way and Local Group is $1\text{-}3\times10^6$ K, and while they both may have the identical temperature, there are reasons to believe they are different. Galactic X-ray emission studies indicate a temperature of $1.2\times10^6$ K for the thick halo (Snowden et al. 2000). In contrast, a higher temperature is expected for a Local Group medium, or of a medium that was expelled from the Galaxy ($2\text{-}3\times10^6$ K, if not greater). In this temperature range, differences of only a factor of two lead to a dramatically different set of C, N, and O column densities, as inferred from Figure 2. By combining emission observations in the same direction as absorption line data, we can infer information about the abundance, if the gas is cospatial. The emission measure contains the electron density while the absorption is proportional to the ionic column, so by combining them, abundances can be derived.

A complementary diagnostic for distinguishing between Galactic and Local Group components is the velocity of the gas, which should vary by a few hundred km $s^{-1}$. For example, the difference between the heliocentric velocity and the Galactocentric velocity can be up to 316 km $s^{-1}$, depending on the line-of-sight. If there is a hot Galactic halo that is rotating with the Milky Way, it will produce a clear signature of velocity as a function of longitude. Relative to the Galactocentric velocity, M31 differs by $-250$ km $s^{-1}$ and the barycenter of the Local Group differs by about $-100$ km $s^{-1}$. The velocity differences depend on latitude and longitude on the sky, so with good velocity centroids, the different components are separable. With sufficiently high velocity resolution, one could separate components that have similar ionic distributions. The intrinsic Doppler widths of the lines are about 50 km $s^{-1}$ and the sound speed of the gas is 100 km $s^{-1}$ (typical of turbulence, if present), so component identification is possible in principle, but impossible at the resolution of existing instruments (700 km $s^{-1}$).

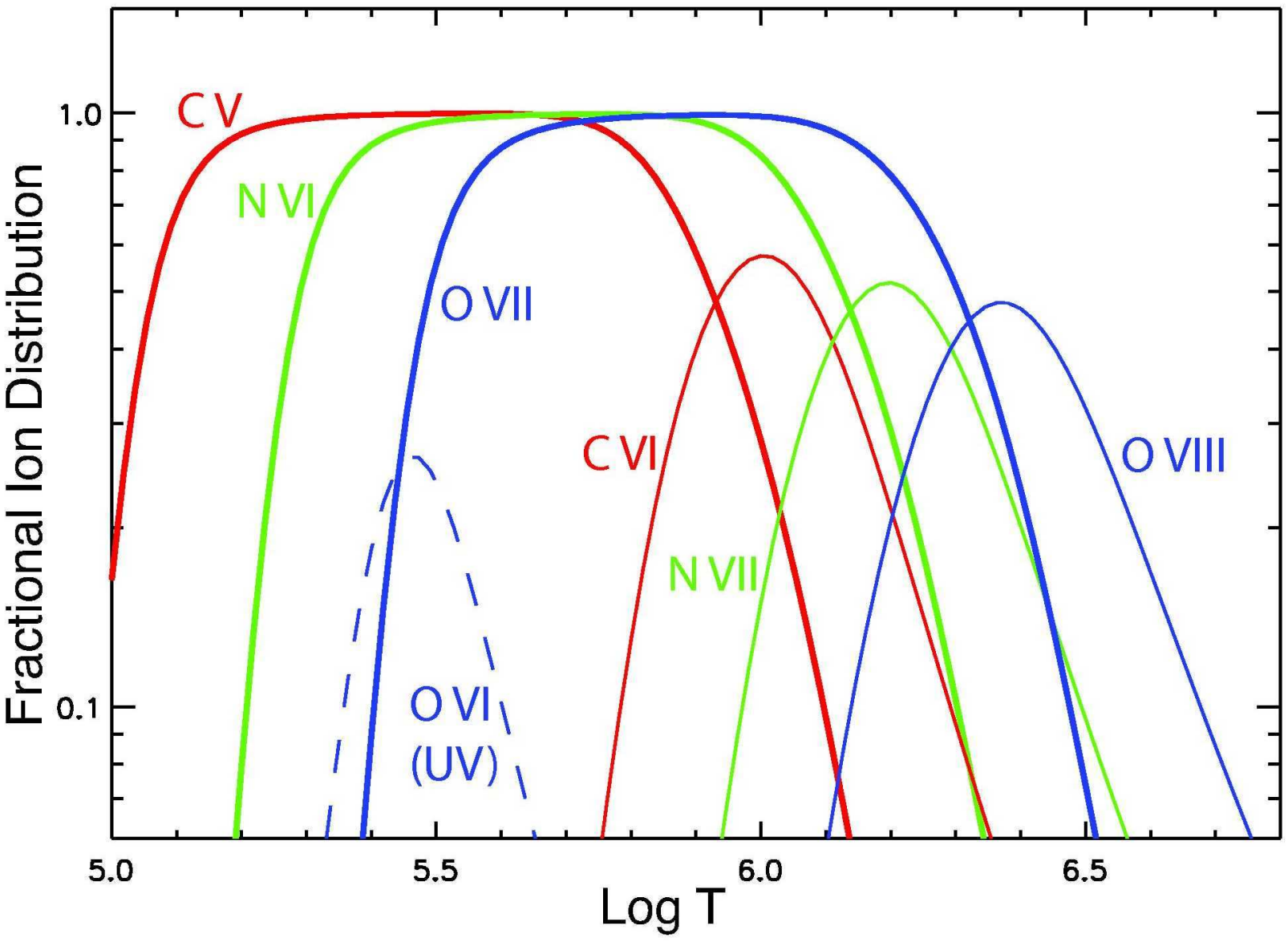

**Figure 2.** The fractional ionization distribution for the six most easily detectable ions from X-ray absorption line measurements, plus OVI (a UV line). The ions C VI, N VII, and O VIII should trace the ambient gas in the potential well of the Milky Way and Local Group, which has an equivalent temperature of 1-3×$10^6$ K. The lower temperature ions, such as O VI and C V, trace cooling gas or material in conductive or turbulent interfaces. The presence of multiple elements at similar temperatures allows one to measure abundance ratios.

A third and very powerful approach for separating the various components is the angular distribution, which was used with *ROSAT* emission observations to reveal the flattened hot gas halo around the Milky Way (Pietz et al. 1998; Snowden et al. 2000). Also, it is possible to separate absorption by hot gas from the Milky Way with that of the Local Group. The Local Group is non-spherical so gas will be extended along the long axis of the system (toward M31; $l$ = 121°) while Galactic gas has its highest column density in sight lines across the Galaxy (Figure 3). Furthermore, the size of a Galactic halo can be inferred from the distribution of column densities with Galactic latitude and longitude. An initial effort along these lines was carried out with *XMM* observations of the OVII absorption line (Bregman and Lloyd-Davies 2007). It supports the presence of a hot Galactic halo, but the size and distribution of the hot halo are poorly known, primarily due to the limited number of sightlines and the poor S/N of the observations. The *XMM* data showed the weaker O VIII resonance line along four sightlines and so was not useful for angular studies.

The gas masses in an extended Galactic halo and in the Local Group are considerable. Current observations give us the ionic column densities, which can be converted to masses when a radius for the absorbing region is adopted. A Galactic halo of radius 100 kpc corresponds to a gas mass of $10^{10.5}$ $M_\odot$, while if the OVII absorption line is caused by Local Group gas, the mass is $10^{12.5}$ $M_\odot$, exceeding the baryon content of all Local Group galaxies(Rasmussen, Kahn, and Paerels 2003; Nicastro et al. 2002); this could account for the missing baryons.

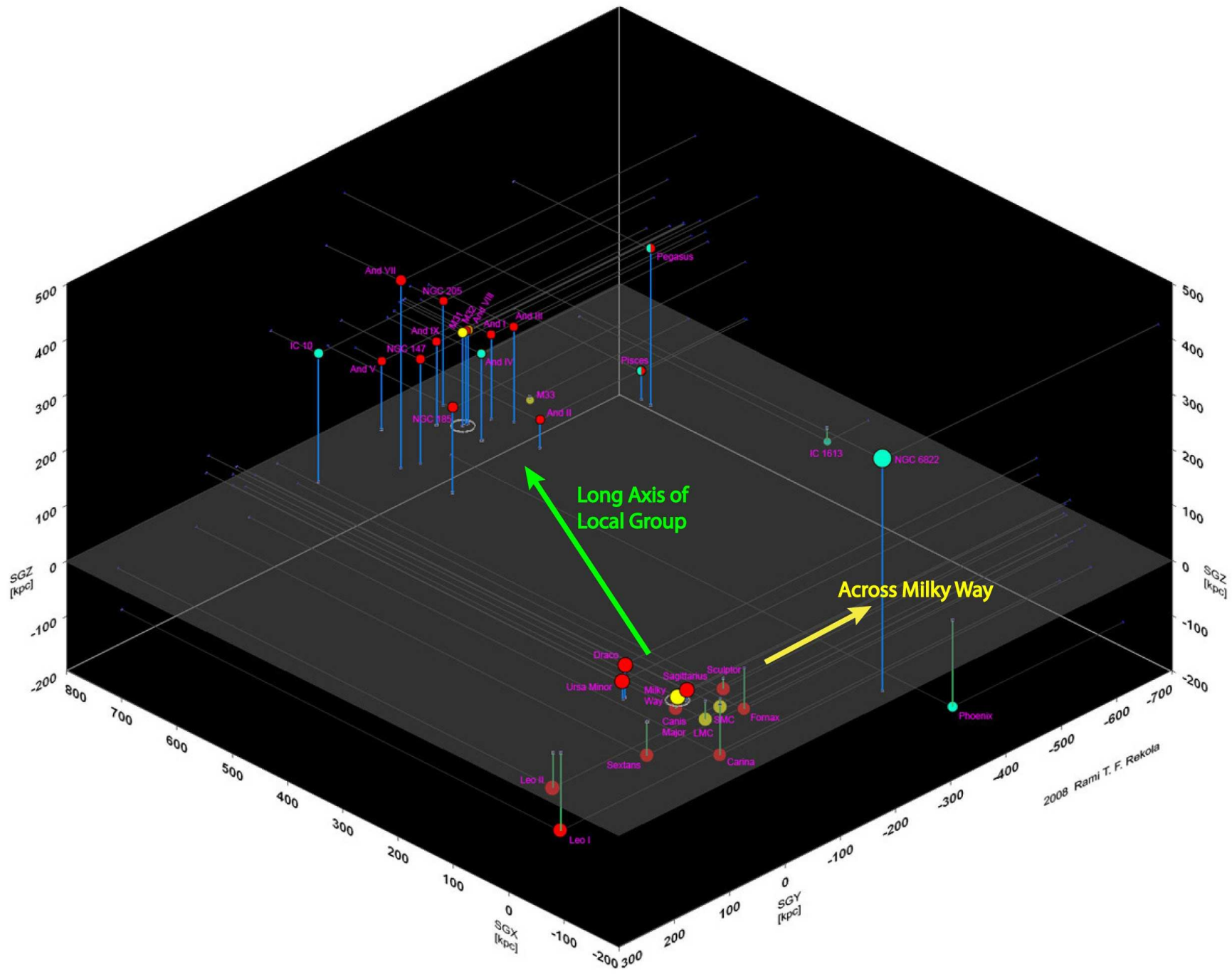

**Figure 3.** A three-dimensional representation of most of the Local Group, showing the vector directions across the Milky Way and along the long axis of the Local group, which differ by 120°.

## 2. Observational Considerations

The observed X-ray absorption lines are 5-20 times larger than those expected from the WHIM, which originate in lower density regions. This is why OVII at z = 0 has been detected along 10 lines of sight at S/N = 3-13 with *XMM* and *Chandra*. Many more lines of sight are needed to determine the size of a Galactic halo and a Local Group medium. To achieve the science goals, we need line centroids of 100 km/sec or less, and column density measurements for at least 100 objects.

Both goals are easily achieved with a future X-ray mission such as the *International X-Ray Observatory* (*IXO*), which has a baseline spectral resolution of 100 km $s^{-1}$ (centroids to 10-30 km/sec can be obtained) and it is 15 times faster than *XMM* for spectral line detection, so we can measure absorption lines toward hundreds of AGNs. Many of these observations will be obtained for free because the gratings are always present, so adequate spectra will be obtained for any observation of an AGN brighter than about $1\times10^{-11}$ erg $cm^{-2}$ $s^{-1}$ that is observed for longer than 30 ksec. More than 100 AGNs will be observed in this parallel mode, based on the usage of previous X-ray telescopes and on other anticipated *IXO* projects. There are many known objects yet brighter, and in the *ROSAT* All Sky Survey, there are 421 objects with fluxes above $1\times10^{-11}$ erg cm$^{-2}$ s$^{-1}$ (most of which are AGN; Figure 4), for which high-quality spectra could be obtained with modest observing times (Figure 5). Investigators might propose targeted programs such as: additional lines of sight across the Milky Way ($-45° < l < 45°$), to better constrain Galactic halo models and measure halo rotation; observations of the hot halos of Local Group galaxies, using AGNs projected within a virial radius; or observations of the Magellanic Stream and high velocity clouds, complementing the UV studies of OVI (Wakker et al. 2003). We expect this to be a vibrant field with IXO that will use several Msec of time, depending on how projects are conducted.

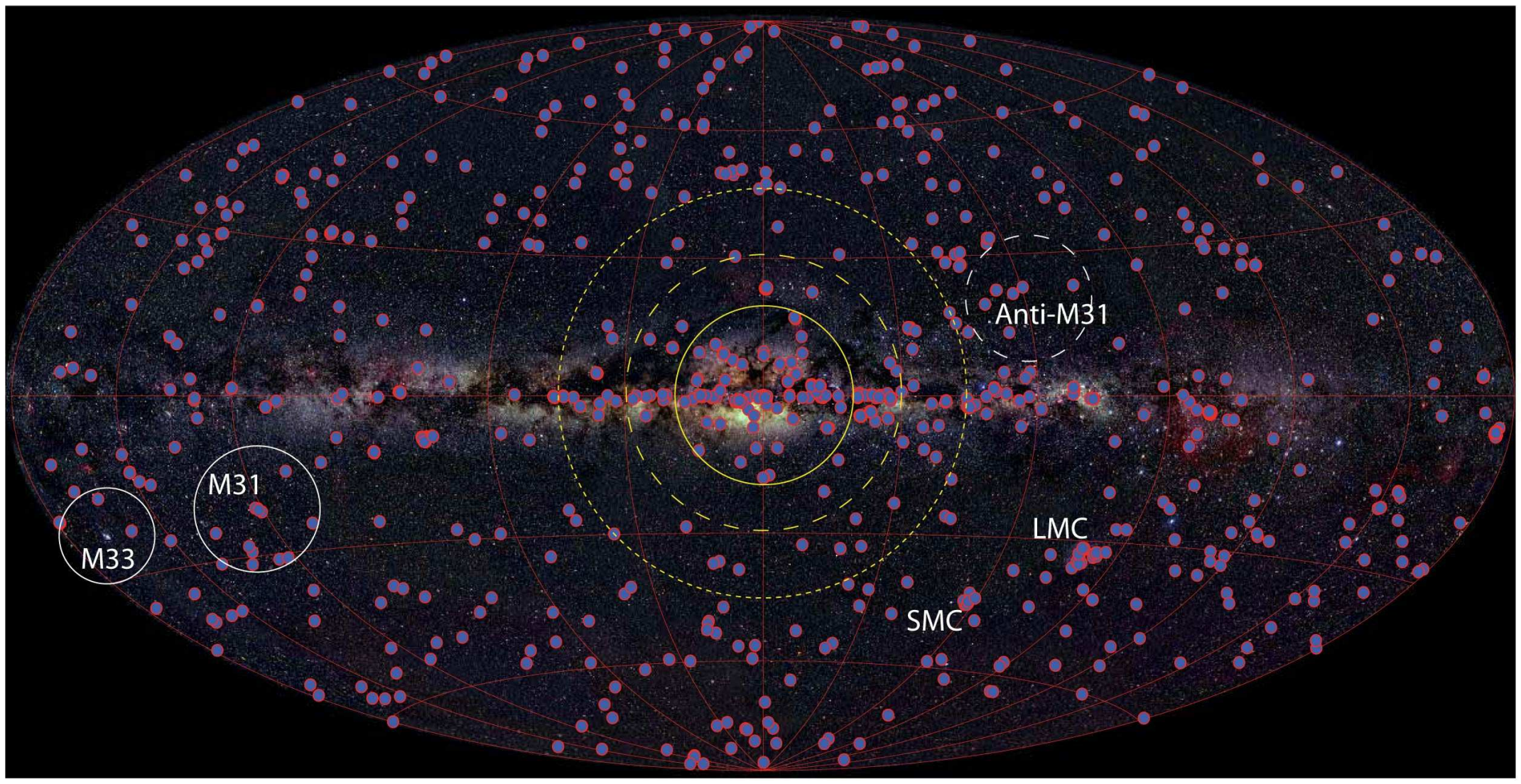

**Figure 4.** An optical image of the Milky way with the 421 X-ray sources with $F_X > 1\times10^{-11}$ erg cm$^{-2}$ s$^{-1}$ (0.5-2.5 keV), all of which can produce high-quality X-ray absorption line spectra (Fig. 5; nearly all sources away from the plane are AGNs). Regions around the bulge (yellow circles) will have the largest columns in a Galactic model and should show redshifts (1$^{st}$ quadrant) and blueshifts (4$^{th}$ quadrant) caused by rotation. A Local Group component should have relatively low column densities in the anti-M31 direction and the greatest columns along the long axis of the Local Group, toward M31 and M33.

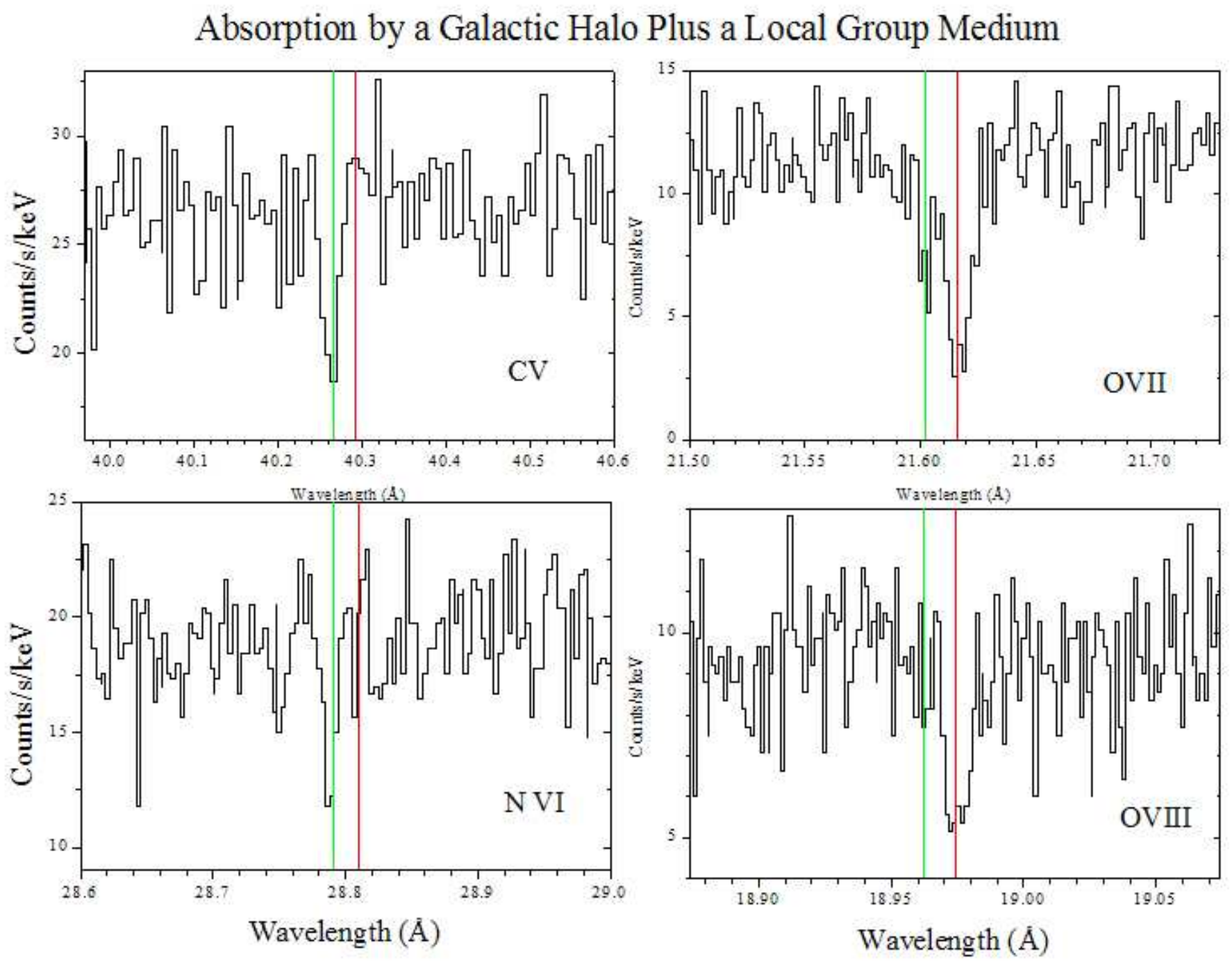

**Figure 5.** A simulated IXO grating spectrum with the baseline resolution and collecting area (R = 3000, 1000 $cm^2$) of a background AGN with the product of the flux and exposure time of ($F_x/2\times10^{-11}$ $erg/cm^2/s$) (t/50 ksec). There are two components in the model, a $0.8\times10^6$ K Galactic halo component (green line) and a $2\times10^6$ K Local Group medium, offset by 200 km/s (red line). The cooler Galactic component is detected at the lower temperature ions while the hotter Local Group medium is detected in the higher temperature ions and shifted in velocity.